\documentclass[twocolumn,english,aps,prl]{revtex4-1}
\usepackage[T1]{fontenc}
\usepackage[latin9]{inputenc}
\usepackage[a4paper]{geometry}
\usepackage{float}
\usepackage{amsthm}
\usepackage{amsmath}
\usepackage{amssymb}
\usepackage{graphicx}
\usepackage{esint}
\usepackage{babel}
\begin{document}

\title{Singularities in FLRW Spacetimes}

\author{Huibert het Lam and Tomislav Prokopec }

\address{Institute for Theoretical Physics, Spinoza Institute and $EMME\Phi$,
Utrecht University,}

\address{Postbus 80.195, 3508 TD Utrecht, The Netherlands}
\begin{abstract}
We point out that past-incompleteness of geodesics in FLRW spacetimes
does not necessarily imply that these spacetimes start from a singularity.
Namely, if a test particle that follows such a trajectory has a non-vanishing
velocity, its energy was super-Planckian at some time in the past
if it kept following that geodesic. That indicates a breakdown of
the particle's description, which is why we should not consider those
trajectories for the definition of an initial singularity. When one
only considers test particles that do not have this breakdown of their
trajectory, it turns out that the only singular FLRW spacetimes are
the ones that have a scale parameter that vanishes at some initial
time. 
\end{abstract}
\maketitle

\section{Introduction}

Hubble's law, the observed abundance of elements, the cosmic background
radiation and the large scale structure formation in the universe
are strong evidence that the universe expanded from an initial very
high dense state to how we observe it now. However, what happened
exactly during this hot density state is still an open problem. One
of the questions that needs to be answered is whether there was a
singularity at the beginning of spacetime. Such a singularity is in
accordance with the very general theorems of Hawking and Penrose \cite{hawking1973large},
\cite{Hawking:1969sw} defined as a non-spacelike geodesic that is
incomplete in the past. One uses this definition because test particles
move on these trajectories and thus have only traveled for a finite
proper time.

The flatness, horizon and magnetic monopole problem can be solved
with a period of exponential expansion in the very early universe
\cite{Starobinsky:1980te}, \cite{Guth:1980zm}. To avoid a singularity
before that period, it was suggested that one can have past-eternal
inflation in which the universe starts from an almost static universe
and flows towards a period of exponential expansion. This way the
universe would not have a beginning. One of the characteristics of
inflationary models is that the Hubble parameter $H$ is positive.
In \cite{Borde:2001nh} it was shown that when the average Hubble
parameter along a geodesic $H_{\mathrm{av}}$ is positive, the geodesic
is past-incomplete such that we would have a singularity. This is
also applicable to models of eternal inflation in which the average
Hubble parameter along geodesics does not go to zero sufficiently
fast (i.e. such that we do not have that $H_{\mathrm{av}}$ is zero).
In \cite{Ellis:2002we}, a model of eternal inflation was given with
all non-spacelike geodesics complete, but in \cite{Mithani:2012ii}
these kind of models were shown to be quantum mechanically unstable.
Hence, this would imply that also models of eternal inflation start
from a singularity. 

In \cite{Aguirre:2001ks} it was pointed out that in De Sitter space
the test particles that follow those past-incomplete trajectories
and have a non-vanishing velocity, will have an energy that becomes
arbitrarily large when going back in the past. This can be generalized
to general Friedmann-Lemaître-Robertson-Walker (FLRW) spacetime and
means that the energy of such a test particle can become super-Planckian
at some initial time such that their description breaks down. This
is the reason one should not consider those trajectories when defining
a singularity. When one only considers the trajectories of test particles
that do not have a breakdown of the description of their trajectory,
one finds that the only FLRW spacetimes that start from a singularity
are the ones with a scale factor that vanishes at some initial time.
This implies that models of eternal inflation or bouncing models are
singularity free provided one requires sub-Planckian test particles
at all times.

In this paper we first consider the past-(in)completeness of geodesics
in spacetimes with an FLRW metric. We review the general singularity
theorems of \cite{hawking1973large}, \cite{Hawking:1969sw} applied
to these models and we review the more general (in the context of
cosmology) argument of \cite{Borde:2001nh}. After that we consider
how the energies of test particles change in time. We adopt units
in which the velocity of light $c=1$.

\section{Past-(in)completeness of Geodesics in FLRW Spacetimes}

Consider a universe with an FLRW metric which describes a spatially
homogeneous, isotropic spacetime: 
\begin{equation}
ds^{2}=-dt^{2}+a(t)^{2}\left[\frac{dr^{2}}{1-\kappa r^{2}}+r^{2}\left(d\theta^{2}+\sin^{2}(\theta)d\varphi^{2}\right)\right],\label{eq:flrw metric}
\end{equation}
where $\kappa$ is the curvature of spacelike three-surfaces and the
scale factor $a(t)$ is normalized such that $a(t_{1})=1$ for some
time $t_{1}$. This metric is a good description of our universe,
since from experiments as WMAP and Planck, it follows that our universe
is spatially homogeneous and isotropic when averaged over large scales.
Geodesics $\gamma(\tau)$, where $\tau$ is an affine parameter, satisfy
\begin{equation}
\frac{d\gamma^{0}}{d\tau}=\frac{\sqrt{|\vec{V}(t_{1})|^{2}-\epsilon a^{2}}}{a},\label{eq:time component geodesic}
\end{equation}
where $|\vec{V}|^{2}=g_{ij}\dot{\gamma}^{i}\dot{\gamma}^{j}$ and
$\epsilon$ is the normalization of the geodesic: $\epsilon=0$ for
null geodesics and $\epsilon=-1$ for timelike geodesics. We thus
have a past-incomplete geodesic when 
\begin{equation}
\int d\tau=\int_{t_{0}}^{t}\frac{a}{\sqrt{|\vec{V}(t_{1})|^{2}-\epsilon a^{2}}}dt\label{eq:condition incomplete geodesic}
\end{equation}
for an initial velocity $|\vec{V}(t_{1})|$ is finite. Here $t_{0}$
is $-\infty$ if $a(t)>0$ for all $t$, otherwise $t_{0}\in\mathbb{R}$
is taken such that $a(t_{0})=0$. Notice that when $a(t_{0})=0$ for
some time $t_{0}$, all non-spacelike geodesics are past-incomplete.
When $t_{0}=-\infty$ and the integral (\ref{eq:condition incomplete geodesic})
is converging, we cannot immediately conclude that geodesics are past-incomplete.
It is possible that we only consider a part of the actual spacetime.
An example is given by $\kappa=0,$ and the Hubble parameter $H=\dot{a}/a$
satisfying $\dot{H}/H^{2}=0,$ in which case $a(t)=e^{Ht}$ with $H$
constant. If the whole manifold would be covered by these coordinates,
it would result in past-incomplete geodesics. However, this model
only describes one half, known as the Poincaré patch, of the larger
De Sitter space; the whole space is described by choosing $\kappa=1,$
$a(t)=\cosh(Ht)/H$ which yields complete geodesics. See also \cite{Aguirre:2003ck}
and \cite{Aguirre:2007gy}. When the integral (\ref{eq:condition incomplete geodesic})
is diverging one can conclude that geodesics in that specific coordinate
patch are past-complete. Of course, one can also assume that a certain
model with $t_{0}=-\infty$ covers the whole spacetime. Then the past-(in)completeness
of a geodesic is determined by the integral (\ref{eq:condition incomplete geodesic}).

From (\ref{eq:condition incomplete geodesic}) we see that in spacetimes
with $a(t)>A\in\mathbb{R}_{>0}$ all non-spacelike geodesics are past-complete.
Hence for a spacetime to have a non-spacelike geodesic that is past-incomplete,
$a(t)$ needs to become arbitrarily small. 

\begin{figure}[H]
\begin{centering}
\includegraphics[scale=0.6]{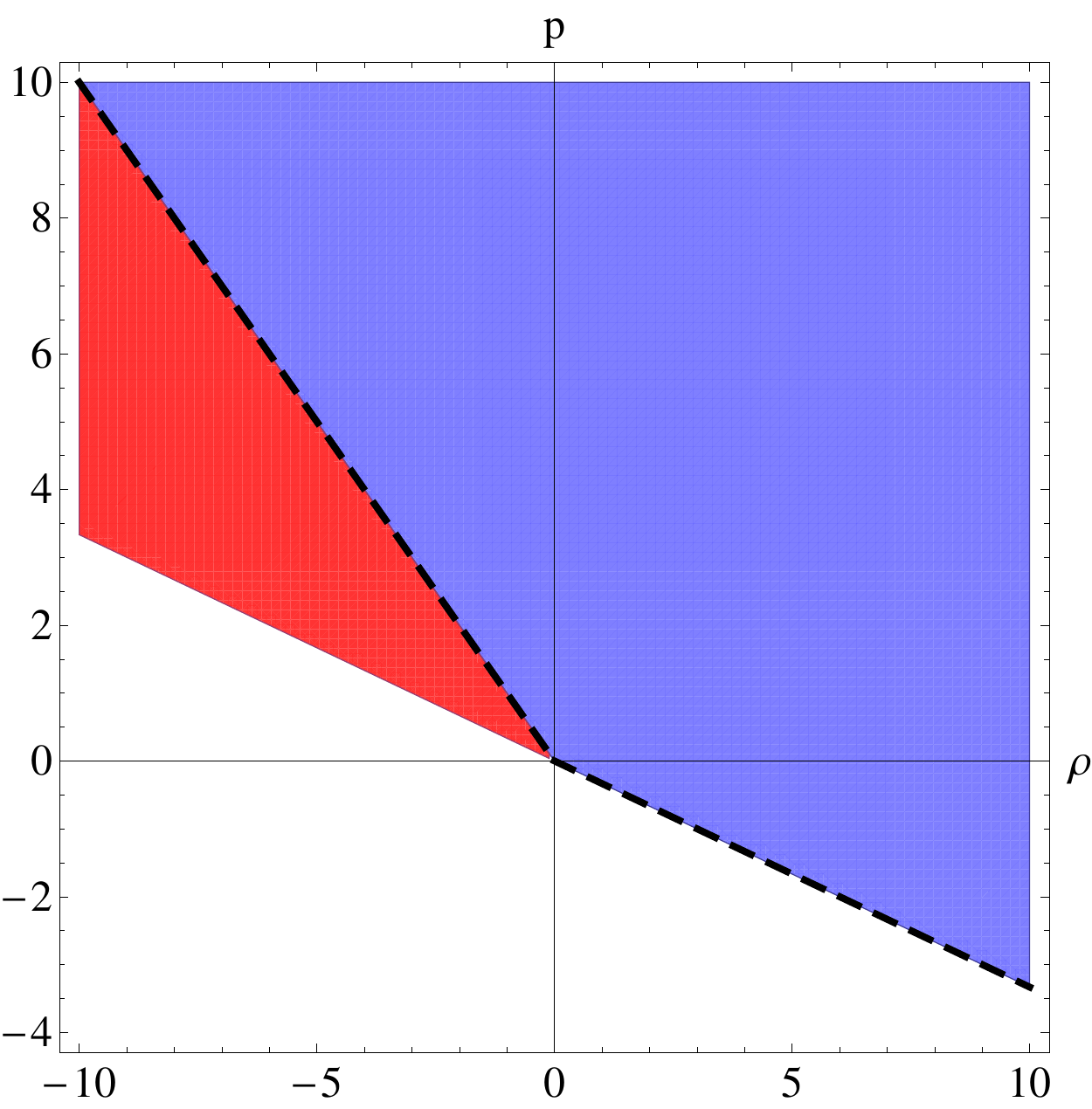}
\par\end{centering}

\begin{centering}
\caption{\label{fig:Illustration-of-condition}Illustration of condition (\ref{eq:physical condition singularity theorems flrw}).
For $\kappa<0$ one needs $(\rho,p)$ in the blue shaded area above
the dashed line to apply the Hawking-Penrose singularity theorems.
For $\kappa\geq0,$ we have less restrictions, the red shaded area
below the dashed line is also included, but it is impossible for an
FLRW spacetime with non-negative spatial curvature to be in that area.}

\par\end{centering}

\end{figure}

There are a few theorems that prove that a spacetime contains a (past-)incomplete
geodesic. Hawking and Penrose, \cite{hawking1973large}, \cite{Hawking:1969sw},
proved theorems that state that when 
\begin{equation}
R_{\mu\nu}\dot{\gamma}^{\mu}\dot{\gamma}^{\nu}\geq0\label{eq:condition singularity theorems}
\end{equation}
for all geodesics $\gamma$ and the spacetime obeys a few other conditions
such as containing a trapped surface, there is a non-spacelike geodesic
that is incomplete. Condition (\ref{eq:condition singularity theorems})
for the metric (\ref{eq:flrw metric}) yields
\begin{equation}
\left(\frac{\overset{..}{a}}{a}+2\frac{\dot{a}^{2}}{a^{2}}+2\frac{\kappa}{a^{2}}\right)\epsilon-2\left[\frac{\overset{..}{a}}{a}-\frac{\dot{a}^{2}}{a^{2}}-\frac{\kappa}{a^{2}}\right]\left(\dot{\gamma}^{0}\right)^{2}\geq0.\label{eq:condition singularity theorems frlw}
\end{equation}
Using Eq. (\ref{eq:time component geodesic}) one finds that condition
(\ref{eq:condition singularity theorems frlw}) becomes
\begin{eqnarray}
\kappa\geq0 & : & \overset{..}{a}\leq0;\nonumber \\
\kappa<0 & : & \begin{cases}
\frac{\overset{..}{a}}{a}-\frac{\dot{a}^{2}}{a^{2}}-\frac{\kappa}{a^{2}} & \leq0;\\
\overset{..}{a} & \leq0.
\end{cases}\label{eq:geometric singularity conditions flrw}
\end{eqnarray}
In particular for all $\kappa$ we need that $\overset{..}{a}\leq0$
at all time, or that the spacetime is non-accelerating. Notice that
when $\overset{..}{a}\leq0$, $a$ will always be zero at some time
$t_{0}$ (this might be in the future), unless $a$ is a positive
constant ($H=0$) in which case we do not have past-incomplete geodesics.
Hence, when we want to use these theorems to say something about an
initial singularity in an FLRW spacetime, we need a metric that has
a scale parameter $a$ that becomes zero at some time in the past.
Describing the matter content of the universe by a perfect fluid
\begin{equation}
T_{\mu\nu}=(\rho+p)U_{\mu}U_{\nu}+pg_{\mu\nu},
\end{equation}
where $p$ is the pressure, $\rho$ the energy density and $U^{\mu}=(1,0,0,0)$,
the condition (\ref{eq:geometric singularity conditions flrw}) translates
via the Friedmann equations to
\begin{eqnarray}
\kappa\geq0 & : & \rho+3p\geq0;\nonumber \\
\kappa<0 & : & \begin{cases}
\rho+p & \geq0;\\
\rho+3p & \geq0.
\end{cases}\label{eq:physical condition singularity theorems flrw}
\end{eqnarray}
Although it seems that we have less restrictions when $\kappa\geq0,$
it is impossible that $\rho+p<0$ and $\rho+3p\geq0$ for non-negative
spatial curvature. In Fig. \ref{fig:Illustration-of-condition} one
finds an illustration of condition (\ref{eq:physical condition singularity theorems flrw}).

Another theorem that proves that a geodesic is past-incomplete was
published in \cite{Borde:2001nh} and is also applicable to spacetimes
that have $a(t)>0$ for all $t$. It says that when the average Hubble
parameter $H=\dot{a}/a$ along a non-spacelike geodesic, $H_{\mathrm{av}}$,
satisfies $H_{\mathrm{av}}>0$, the geodesic must be past-incomplete.
For the metric (\ref{eq:flrw metric}), the argument is as follows.
Consider a non-spacelike geodesic $\gamma(\tau)$ between an initial
point $\gamma(\tau_{\mathrm{i}})$ and a final point $\gamma(\tau_{\mathrm{f}})$.
We can integrate $H$ along the geodesic, using Eq. (\ref{eq:time component geodesic}):
\begin{eqnarray}
\int_{\tau_{\mathrm{i}}}^{\tau_{\mathrm{f}}}Hd\tau & = & \int_{t_{\mathrm{i}}}^{t_{\mathrm{f}}}\frac{\dot{a}}{\sqrt{|\vec{V}(t_{1})|^{2}-\epsilon a^{2}}}dt\nonumber \\
 & = & \int_{a(t_{\mathrm{i}})}^{a(t_{\mathrm{f}})}\frac{da}{\sqrt{|\vec{V}(t_{1})|^{2}-\epsilon a^{2}}}\\
 & = & \begin{cases}
\frac{1}{|\vec{V}(t_{1})|}\left[a(t_{\mathrm{f}})-a(t_{\mathrm{i}})\right], & \epsilon=0\\
\log\left(\frac{a(t_{\mathrm{f}})+\sqrt{|\vec{V}(t_{1})|^{2}+a(t_{\mathrm{f}})^{2}}}{a(t_{\mathrm{i}})+\sqrt{|\vec{V}(t_{1})|^{2}+a(t_{\mathrm{i}})^{2}}}\right) & \epsilon=-1
\end{cases}\nonumber \\
 & \leq & \begin{cases}
\frac{a(t_{\mathrm{f}})}{|\vec{V}(t_{1})|}, & \epsilon=0\\
\log\left(\frac{a(t_{\mathrm{f}})+\sqrt{|\vec{V}(t_{1})|^{2}+a(t_{\mathrm{f}})^{2}}}{|\vec{V}(t_{1})|}\right) & \epsilon=-1.
\end{cases}\nonumber 
\end{eqnarray}
Notice that for the second equality sign, one should break up the
integration domain into parts where $a=a(t)$ is injective, but that
one will end up with the same result. Hence, this integral as function
of the initial affine parameter $\tau_{\mathrm{i}}$ is restricted
by some fixed final $\tau_{\mathrm{f}}$. This means that when 
\begin{equation}
H_{\mathrm{av}}=\frac{1}{\tau_{\mathrm{f}}-\tau_{\mathrm{i}}}\int_{\tau_{\mathrm{i}}}^{\tau_{\mathrm{f}}}Hd\tau>0
\end{equation}
$\tau_{\mathrm{i}}$ has to be some finite value such that the geodesic
is past-incomplete. Notice that it is still possible to construct
an FLRW spacetime that has $H>0$ at all times and complete geodesics.
For this we need that $H_{\mathrm{av}}$ must become zero when $\tau_{\mathrm{i}}\rightarrow-\infty$.
Examples are for instance given by spacetimes with $H>0$ and $a\rightarrow a_{0}>0$
for $t\rightarrow-\infty$ (in this case we will have that $H\rightarrow0$
as $t\rightarrow-\infty$).

\section{Energy of Test Particles}

As stated before, the definition of a singularity is based on the
trajectories of massive test particles and massless particles. For
cosmological spacetimes with an FLRW metric, we would like to study
the energies of test particles over time. We will generalize the argument
given in \cite{Aguirre:2001ks} for De Sitter space to a general FLRW
spacetime. 

Using Eq. (\ref{eq:time component geodesic}) we find that for massive
test particles
\begin{eqnarray}
|\vec{V}|^{2} & = & g_{ij}\dot{\gamma}^{i}\dot{\gamma}^{j}=\epsilon+\left(\dot{\gamma}^{0}\right)^{2}=\frac{|\vec{V}(t_{1})|^{2}}{a^{2}}.\label{eq:velocity testparticle}
\end{eqnarray}
We already saw that in order for a spacetime to have a past-incomplete
non-spacelike geodesic, the scale parameter $a$ needs to become arbitrarily
small. With Eq. (\ref{eq:velocity testparticle}) this then implies
that when the particle has a velocity $|\vec{V}(t_{1})|$ at time
$t_{1},$ the velocity and hence the energy $E^{2}=m^{2}\left(1+\frac{|\vec{V}(t_{1})|^{2}}{a^{2}}\right)$
of a test particle with mass $m$ become arbitrarily large when moving
back to the past. 

The statement above for massive test particles carries over to photons.
In this case the angular frequency as observed by a comoving observer
is
\begin{equation}
\omega=\dot{\gamma}^{0}=\frac{\omega(t_{1})}{a}.
\end{equation}
Thus also the energy of photons $E=\hbar\omega$ will become arbitrarily
large when moving back to the past.

In \cite{Aguirre:2001ks} it was noted that one cannot have particles
with arbitrarily high energies because if such a particle has a nonvanishing
interaction cross section with any particle with a non-zero physical
number density, then the particle will interact with an infinite number
of them, breaking the Cosmological principle. However, the particle's
energy cannot become arbitrarily high because it will reach the Planck
energy $E_{\mathrm{P}}=\sqrt{\frac{\hbar}{G}}\approx1.22\cdot10^{19}\mathrm{GeV}$
at some time $t$. With this energy, the particle's Compton wavelength
is approximately equal to its Schwarzschild radius such that it will
form a black hole. Therefore, the description of the particle's trajectory
will break down. Scattering processes involving vacuum fluctuations
may cause the test particle's energy to never reach the Planck energy.
If these processes are significant the particle's trajectory is not
a geodesic anymore. Near the Planck energy scattering processes are
dominated by processes that involve the exchange of a graviton \cite{'tHooft:1987rb}.
To estimate this effect we consider photon-photon scattering with
the exchange of a graviton. We model the loss of energy of the photon
when going back in time as 
\begin{equation}
\frac{d}{dt}E=\left(-H-\sigma n\right)E,\label{eq:increase energy}
\end{equation}
where $n$ is the number density of virtual photons and $\sigma$
is the cross section of the scattering process. The particle gains
energy from the expansion of the universe because $-H$ is positive
(when going back in time) and it looses energy from the scattering
with virtual photons. We estimate the density of virtual photons as
one per Hubble volume: 
\begin{equation}
n=\frac{1}{V_{\mathrm{H}}}=-\frac{3H^{3}}{4\pi}.\label{eq:number density}
\end{equation}
The differential cross section for photon-photon scattering with the
exchange of a graviton for unpolarized photons is \cite{Boccaletti:1969aj}
\begin{eqnarray}
\frac{d\sigma}{d\Omega} & = & \frac{\kappa^{4}}{8\pi^{2}}\frac{k^{2}}{\sin^{2}(\theta)}\left[1+\cos^{16}\left(\frac{1}{2}\theta\right)+\sin^{16}\left(\frac{1}{2}\theta\right)\right]\nonumber \\
 &  & \mbox{}
\end{eqnarray}
where $\kappa=\sqrt{16\pi G}$, $k$ is the momentum of the photon
and $\theta$ is the scattering angle. Since we are primarily interested
in large momentum exchange, we neglect small angle scatterings when
calculating the total cross section of this process: 
\begin{eqnarray}
\sigma & = & \int\frac{d\sigma}{d\Omega}d\Omega\nonumber \\
 & = & \frac{\kappa^{4}}{\pi}\frac{k^{2}}{4}\int_{-1+\xi}^{1-\xi}\frac{1+\frac{1}{256}\left(1+x\right)^{8}+\frac{1}{256}\left(1-x\right)^{8}}{1-x^{2}}dx\nonumber \\
 & = & \frac{\kappa^{4}}{\pi}\frac{k^{2}}{2}\int_{\xi}^{1}\frac{1+\frac{1}{256}\left(2-y\right)^{8}+\frac{1}{256}y^{8}}{y(2-y)}dy\nonumber \\
 & = & \frac{\kappa^{4}}{\pi}\frac{k^{2}}{4}\left[2\log\frac{1}{\xi}-\frac{363}{140}+\log(4)+\mathcal{O}(\xi)\right],\label{eq:cross section}
\end{eqnarray}
where we have the relation $\mbox{\ensuremath{\sin}(\ensuremath{\theta}/2)=\ensuremath{\sqrt{\xi/2}}}$.
Taking only angles $.26\pi<\theta<.74\pi$ into account for the scattering,
we have that $2\log\frac{1}{\xi}-\frac{363}{140}+\log(4)\approx1.$
With Eqs. (\ref{eq:increase energy}), (\ref{eq:number density})
and (\ref{eq:cross section}) we find that the energy of the test
photon does not increase when
\begin{equation}
H\sim\sigma n=48G^{2}E^{2}H^{3},\label{eq:condition}
\end{equation}
where $E=k$ is the photon energy. Using the Hubble parameter of cosmic
inflation which typically is about $-\hbar H\approx10^{13}\;\mathrm{GeV}$,
we find from (\ref{eq:condition}) that the scattering process becomes
significant when 
\begin{equation}
\left(\frac{E}{E_{\mathrm{P}}}\right)^{2}\sim\frac{E_{\mathrm{P}}^{2}}{48\hbar^{2}H^{2}}\approx10^{10}.
\end{equation}
Hence, processes involving gravitons will not cause the particle's
energy to stay smaller than the Planck energy and a black hole will
form. This implies that the description of the particle's trajectory
(as a geodesic) breaks down, either because of interaction processes
or by the formation of a black hole. The latter definitely happens
when the initial energy is near the Planck energy.

Up to now, the maximum energy of a single particle that has been measured
is of the order of $10^{20}\;\mathrm{eV}$ \cite{ThePierreAuger:2015rha}
which is eight orders of magnitude smaller than the Planck scale.
These particles were all cosmic ray particles, so their probable origin
is a supernova, an active galactic nucleus, a quasar or a gamma ray-burst.
Even when using this energy as an upper bound for the energy of test
particles, we have that the description of the trajectories of non-commoving
test particles breaks down at times that are certainly later than
the Planck era, the period where we have to take quantum gravitational
effects into account. In \cite{Aguirre:2001ks} the arbitrarily high
energies of test particles were used to argue that these particles
should be forbidden in De Sitter space. This can be done by using
a different time arrow in the two patches of De Sitter space that
one has in the flat slicing. That way the two coordinate patches become
non-communicating and describe eternally inflating spacetimes. We
will not look into these kind of constructions for general FLRW spacetimes
but we want to use the arbitrarily high energies of test particles
to give a consistent definition of a singularity. When the particle's
description breaks down before it reaches the beginning of its trajectory,
it is not very useful to use that particle as an indication for an
initial singularity. That is the reason why we suggest to define a
singularity in spacetimes with an FLRW metric that has a parameter
$a$ that becomes arbitrarily small, as a timelike geodesic with $|\vec{V}(t_{1})|=0$
that is past-incomplete. For such trajectories, we have that $dt=d\tau$
which means that a spacetime has no initial singularity when $a(t)>0$
for all $t\in\mathbb{R}$. Hence, an FLRW spacetime starts from a
singularity precisely when $a(t_{0})=0$ at some initial finite time
$t_{0}$.

\section{Conclusion}

We pointed out that spacetimes with an FLRW metric such that $a(t)>0$
for all $t\in\mathbb{R}$ have no initial singularity. This was done
by first observing that in models that have $a(t)>A\in\mathbb{R}_{>0}$
all non-spacelike geodesics are past-complete. When $a$ becomes arbitrarily
small, it is possible that the spacetime contains a past-incomplete
geodesic. With the usual definition of a singularity, this means that
the spacetime has an initial singularity. However, that definition
is based on a test particle that has that geodesic as trajectory.
We pointed out that when this particle has an initial velocity, its
energy will become super-Planckian at some time in the past if it
kept following that geodesic. This means that the particle stops being
a test particle and it does not matter that its trajectory is past-incomplete.
For a model in which the scale factor becomes arbitrarily small, we
should define an initial singularity as a trajectory of a comoving
particle that is past-incomplete. This implies that the only FLRW
spacetimes with an initial singularity are the ones such that $a(t_{0})=0$
at some initial time $t_{0}$. Hence, bouncing spacetimes and past-eternal
inflationary models do not start from a singularity. One can use similar
arguments to show that the only FLRW spacetimes that have a singularity
in the future are the ones that have a scale factor such that $a(t)$
vanishes at some time in the future. It would be interesting to examine
if similar results hold for universes that are obtained by perturbating
an FLRW spacetime.

\bibliographystyle{elsarticle-num}
\bibliography{paper}

\begin{thebibliography}{10}
\expandafter\ifx\csname url\endcsname\relax
  \def\url#1{\texttt{#1}}\fi
\expandafter\ifx\csname urlprefix\endcsname\relax\def\urlprefix{URL }\fi
\expandafter\ifx\csname href\endcsname\relax
  \def\href#1#2{#2} \def\path#1{#1}\fi

\bibitem{hawking1973large}
S.~Hawking, G.~Ellis, {The Large Scale Structure of Space-Time}, Cambridge
  Monographs on Mathematical Physics, Cambridge University Press, 1973.

\bibitem{Hawking:1969sw}
S.~W. Hawking, R.~Penrose, {The Singularities of gravitational collapse and
  cosmology}, Proc. Roy. Soc. Lond. A314 (1970) 529--548.

\bibitem{Starobinsky:1980te}
A.~A. Starobinsky, {A New Type of Isotropic Cosmological Models Without
  Singularity}, Phys. Lett. B91 (1980) 99--102.

\bibitem{Guth:1980zm}
A.~H. Guth, {The Inflationary Universe: A Possible Solution to the Horizon and
  Flatness Problems}, Phys. Rev. D23 (1981) 347--356.

\bibitem{Borde:2001nh}
A.~Borde, A.~H. Guth, A.~Vilenkin, {Inflationary space-times are incomplete in
  past directions}, Phys. Rev. Lett. 90 (2003) 151301.
\newblock \href {http://arxiv.org/abs/gr-qc/0110012}
  {\path{arXiv:gr-qc/0110012}}.

\bibitem{Ellis:2002we}
G.~F.~R. Ellis, R.~Maartens, {The emergent universe: Inflationary cosmology
  with no singularity}, Class. Quant. Grav. 21 (2004) 223--232.
\newblock \href {http://arxiv.org/abs/10211082} {\path{arXiv:10211082}}.

\bibitem{Mithani:2012ii}
A.~Mithani, A.~Vilenkin, {Did the universe have a beginning?}\href
  {http://arxiv.org/abs/1204.4658} {\path{arXiv:1204.4658}}.

\bibitem{Aguirre:2001ks}
A.~Aguirre, S.~Gratton, {Steady state eternal inflation}, Phys. Rev. D65 (2002)
  083507.
\newblock \href {http://arxiv.org/abs/astro-ph/0111191}
  {\path{arXiv:astro-ph/0111191}}.

\bibitem{Aguirre:2003ck}
A.~Aguirre, S.~Gratton, {Inflation without a beginning: A Null boundary
  proposal}, Phys. Rev. D67 (2003) 083515.
\newblock \href {http://arxiv.org/abs/gr-qc/0301042}
  {\path{arXiv:gr-qc/0301042}}.

\bibitem{Aguirre:2007gy}
A.~Aguirre, {Eternal Inflation, past and future}\href
  {http://arxiv.org/abs/0712.0571} {\path{arXiv:0712.0571}}.

\bibitem{'tHooft:1987rb}
G.~'t~Hooft, {Graviton Dominance in Ultrahigh-Energy Scattering}, Phys. Lett.
  B198 (1987) 61--63.

\bibitem{Boccaletti:1969aj}
D.~Boccaletti, V.~De~Sabbata, P.~Fortini, C.~Gualdi, {Photon-photon scattering
  and photon-scalar particle scattering via gravitational interaction
  (one-graviton exchange) and comparison of the processes between classical
  (general-relativistic) theory and the quantum linearized field theory}, Nuovo
  Cim. B60 (1969) 320--330.

\bibitem{ThePierreAuger:2015rha}
A.~Aab, et~al., {Measurement of the cosmic ray spectrum above 4 $\times$
  10$^{18}$ eV using inclined events detected with the Pierre Auger
  Observatory}, JCAP 1508 (2015) 049.
\newblock \href {http://arxiv.org/abs/1503.07786} {\path{arXiv:1503.07786}}.

\end{thebibliography}

\end{document}